\documentclass[a4paper,12pt]{article}

\usepackage[margin=1in]{geometry}
\usepackage{setspace}

\usepackage{float}
\usepackage{amsmath,amssymb}
\usepackage{courier}

\usepackage{graphicx}
\usepackage{tabularx,booktabs,makecell,array,multirow}
\usepackage{siunitx}
\usepackage[round,authoryear]{natbib}

\usepackage{authblk}
\usepackage[hidelinks]{hyperref}

\title{Inferring Transmission Dynamics of Respiratory Syncytial Virus from Houston Wastewater}

\author[1]{Jose R.~Palacio\thanks{Corresponding author: \href{mailto:jrp16@rice.edu}{jrp16@rice.edu}}}
\author[1]{Katherine B.~Ensor}
\author[2]{Sallie A.~Keller}
\author[3]{Rebecca Schneider}
\author[3]{Kaavya Domakonda}
\author[1,3]{Loren Hopkins}
\author[4]{Lauren B.~Stadler}

\affil[1]{Department of Statistics, Rice University, 6100 Main St, Houston, TX, USA}
\affil[2]{Biocomplexity Institute, University of Virginia, Charlottesville, VA, USA}
\affil[3]{Houston Health Department, Houston, TX, USA}
\affil[4]{Department of Civil and Environmental Engineering, Rice University, Houston, TX, USA}

\begin{document}
\maketitle
\begin{abstract}
Wastewater-based epidemiology (WBE) is an effective tool for tracking community circulation of respiratory viruses. We address estimating the effective reproduction number ($R_t$) and the relative number of infections from wastewater viral load. Using weekly Houston data on respiratory syncytial virus (RSV), we implement a parsimonious Bayesian renewal model that links latent infections to measured viral load through biologically motivated generation and shedding kernels. The framework yields estimates of $R_t$ and relative infections, enabling a coherent interpretation of transmission timing and phase. We compare two input strategies—(i) raw viral-load measurements with a log-scale standard deviation, and (ii) state-space–filtered load estimates with time-varying variances—and find no practically meaningful differences in inferred trajectories or peak timing. Given this equivalence, we report the filtered input as a pragmatic default because it embeds week-specific variances while leaving epidemiological conclusions unchanged.
\end{abstract}

\section*{Keywords}
Respiratory Syncytial Virus (RSV), Wastewater-based Epidemiology, Renewal Model, Effective Reproduction Number ($R_t$), Bayesian Inference

\section{Introduction}

Wastewater-based epidemiology (WBE) has established itself as a crucial tool for population-level surveillance of infectious diseases. By quantifying viral genetic material in wastewater, WBE provides a non-invasive and cost-effective monitoring system that can capture early signals of community transmission, independent of medical care-seeking or clinical testing capacity. However, translating these signals into absolute infection counts is challenging due to substantial variability in viral shedding, biological and physical delays in the sewer network, and measurement uncertainty. In practice, it is therefore more informative and robust to focus on relative transmission metrics—most notably the effective reproduction number ($R_t$), which captures temporal changes in transmission dynamics \citep{huisman2022, champredon2024}.

To date, much of the literature has centered on SARS-CoV-2, for which wastewater data have demonstrably supported the public health response. Extending WBE to other pathogens poses additional challenges. For respiratory syncytial virus (RSV)—a leading cause of severe respiratory illness in infants, older adults, and immunocompromised individuals \citep{MayoClinicRSV}—clinical surveillance is affected by underreporting and reporting delays. Many RSV cases are managed at home without hospitalization, so clinical records may not fully reflect community circulation. In wastewater, signal interpretation is further complicated by heterogeneity across treatment plants and the populations they serve \citep{haak2022}, as well as by incubation, shedding dynamics, and transport processes in the sewer system \citep{he2020,hart2020}.

Despite these challenges, RSV is consistently detectable in wastewater, and recent studies highlight its potential as a complement to clinical surveillance \citep{Hughes2021}. Large-scale, systematic analyses in dense urban networks remain scarce, however, leaving open questions about the robustness, comparability, and epidemiological value of wastewater signals \citep{wade2022understanding}.

Multi-stage models are common in environmental epidemiology. In a typical two-stage setup, stage one fits separate models for each unit (e.g., locations or studies) to produce unit-level summaries of the association of interest; stage two then aggregates those summaries to obtain an overall estimate \citep{Sera2022ExtendedTwoStage}. Related Bayesian approaches follow the same spirit: analyze each unit independently, then pass those unit-level posteriors or estimates into a second-stage model that treats them as draws from a common population, allowing for uncertainty propagation \citep{Lunn2013TwoStage}.

In this vein, we adopt a two-stage strategy: outputs from a hierarchical state-space model serve as inputs to a Bayesian model, and we use the resulting posterior to derive downstream epidemiological metrics. The emphasis is on the workflow—separating estimation tasks into modules and carrying forward uncertainty—rather than on any specific implementation detail from prior papers.

Here we outline the structure of the manuscript. Section 2 (Methodology) details the data, the renewal model with generation and shedding kernels, the latent dynamics of $R_t$ and $I_t$, the estimation procedure in \texttt{nimble} \citep{nimble}, and the modular integration with the Kalman filter (SSM). Section 3 (Results) presents convergence diagnostics and parameter estimates, and compares $R_t$ and $I_t$ trajectories across three input/noise specifications (A--C), with an emphasis on the 69th Street plant and sensitivity analyses. Section 4 (Discussion) interprets the key findings---including practical invariance to input choice and stability across unit changes---and their implications for WBE. Finally, Section 5 (Conclusions) summarizes contributions and recommendations.

\section{Statistical Framework and Methodology}

In this work, we analyze weekly RSV viral load expressed in billions of genome copies per day (B gc/day) for a single wastewater treatment plant, the 69th Street WWTP in Houston. Loads are constructed by multiplying observed concentrations (gc/L) by the plant’s median daily influent flow (L/day) and rescaling by $10^{9}$:
\[
y_{t} = C_{t}\,F \times 10^{-9},\qquad t=1,\ldots,T,
\]
where $C_{t}$ is the RSV concentration in week $t$ (gc/L), and $F$ is the median daily influent flow for the wastewater treatment plant (L/day). The resulting $y_t$ denotes viral load in B gc/day. The City of Houston operates 38 facilities serving populations from $>600$ to $>500,000$ residents \citep{houston_publicworks2023}; serving 2.2 million residents. Twenty-four-hour composite samples were collected weekly each Monday between January 16, 2023, and December 30, 2024; values below the laboratory limit of detection (LOD) are treated as missing.

We focus on the 69th Street WWTP, the largest facility in Houston, which serves an estimated 551{,}150 residents. Its size, broad catchment area, and consistent sampling make it one of the city’s most reliable wastewater signals. By concentrating on this plant, we reduce cross-plant heterogeneity and create a clearer setting to test and refine our modeling framework, establishing a methodological benchmark for future extensions across Houston’s network.

We adopt a Bayesian renewal-based framework widely used in wastewater epidemiology  \citep{Fraser2007, huisman2022, champredon2024}. These models link latent infections to wastewater viral loads through two key temporal processes: the generation interval, which governs the renewal of infections, and the shedding profile, which maps past infections to the viral RNA signal.

Whereas pipelines such as \texttt{EpiSewer} adopt a modular structure to incorporate incubation, shedding, and generation dynamics in detail \citep{lison2023}, our version is deliberately less complex. This simplification facilitates interpretation of both the latent transmission process and the plant-specific scaling parameters, while retaining the ability to capture epidemic dynamics from wastewater signals.

We complement the renewal framework with information from a state-space model estimated via the \texttt{MARSS} package \citep{holmes2012marss}. These filtered signals are propagated into the renewal model, allowing us to evaluate robustness and provide an additional perspective on how to evaluate infection dynamics.

Transmission and observation delays are represented by two kernels: the generation interval and the shedding profile. Both are modeled as Gamma distributions in continuous time and discretized into weekly lags using cumulative distribution function differences. This construction is standard in renewal-based epidemic models \citep{Fraser2007, huisman2022, champredon2024}.

The generation interval governs the renewal of infections. Following \citep{Vink2014}, we assume a mean of 7.5 days and a standard deviation of 2.1 days for RSV infections. These values are expressed through a Gamma distribution whose parameters are derived from the reported mean and variance. To obtain weekly weights, we discretize the continuous distribution into intervals centered at integer lags. The support is truncated at $G$ weeks, where $G$ denotes the maximum lag retained in the kernel. Probability mass beyond $G$ is not used, and the retained weights are renormalized to sum to one, following a similar construction to that in \citep{BracherHeld2022}.

For the generation interval, support is therefore $g=1,\dots,G$, excluding lag zero to reflect the biological delay between primary and secondary infections \citep{Fraser2007,huisman2022}:

\begin{equation}
\label{eq:w_gen}
w^{\mathrm{gen}}_{g} \propto
F_{\Gamma}\!\left(g+\tfrac{1}{2};\, \kappa_{\mathrm{GI}}, \theta_{\mathrm{GI}}\right)
-
F_{\Gamma}\!\left(g-\tfrac{1}{2};\, \kappa_{\mathrm{GI}}, \theta_{\mathrm{GI}}\right).
\end{equation}

The shedding profile maps latent infections to RNA viral loads in wastewater.  Based on clinical evidence synthesized by \citep{Cevik2023}, we assume a mean of 4.6 days and a standard deviation of 2.0 days for the duration of viral shedding.  These values are expressed through a Gamma distribution whose parameters are derived from the reported mean and variance.  Because our data are aggregated weekly, the interpretation of lag zero differs from daily formulations: here, $d=0$ corresponds to contributions occurring within the same calendar week as infection, which is biologically plausible given that RSV shedding can begin within the first few days of illness \citep{Hall2001,DeVincenzo2005}. 

As in \eqref{eq:w_gen}, we use a finite support $d=0,\dots,D$; the mass beyond $D$ is omitted and the remaining weights are renormalized to sum to one.

\begin{equation}
\label{eq:w_shed}
w^{\mathrm{shed}}_{d} \propto
F_{\Gamma}\!\left(d+\tfrac{1}{2};\, \kappa_{\mathrm{SH}}, \theta_{\mathrm{SH}}\right)
-
F_{\Gamma}\!\left(\max\{d-\tfrac{1}{2},\,0\};\, \kappa_{\mathrm{SH}}, \theta_{\mathrm{SH}}\right).
\end{equation}

\subsection{Latent Transmission Process}

The effective reproduction number $R_{t}$ is modeled through a scaled softplus transformation of an unconstrained latent link–scale process $z_{t}$ (a unitless transmission-intensity index) \citep{Scott2021epidemia, lison2023}. This transformation guarantees positivity of $R_{t}$ and behaves approximately like the identity in the epidemiologically relevant neighborhood of $R_t \approx 1$, while remaining numerically stable:
\[
R_{t}  =  \frac{\log \bigl(1+\exp(k z_{t})\bigr)}{k}, \qquad k>0.
\]

The parameter $k$ controls the curvature of the link between $z_t$ and $R_t$. Larger values yield a sharper response (approaching a rectified-linear mapping for positive $z_t$), amplifying short-lived increases in transmission, whereas smaller values produce a smoother, more graded mapping. Thus, $k$ regulates how strongly fluctuations in the latent process translate into changes in transmissibility.

The latent process $\{z_{t}\}$ evolves according to a Gaussian random walk, providing a parsimonious yet flexible representation of temporal changes in transmission:
\[
z_{1} \sim \mathcal{N}(1,\sigma_{\epsilon}^{2}), \qquad
z_{t} \sim \mathcal{N}(z_{t-1},\sigma_{\epsilon}^{2}), \quad t \ge 2,
\]
with $\sigma_{\epsilon}>0$ controlling week-to-week volatility.

Let $I_{t}$ denote the number of new infections during week $t$. Following the renewal formulation \citep{Fraser2007}, expected incidence is expressed as a convolution of past infections with the generation-interval distribution:
\[
\lambda_{t}  =  R_{t} \sum_{g=1}^{G} w^{\mathrm{gen}}_{g}  I_{t-g}, \text{ where}
\quad I_{t} \sim \mathrm{Poisson}(\lambda_{t}),  \quad t \ge 2,
\]
and $w^{\mathrm{gen}}_{g}$ are the generation-interval weights normalized to sum to one and $G$ is their maximum support.

Because absolute infections cannot be identified from wastewater alone, the magnitude of \(I_t\) depends on the initial condition, which we set to the citywide average weekly RSV healthcare encounter data multiplied by the share of the city’s population in the service area \citep{HHD_RespiratoryReport}.

\subsection{Viral Load Process}

We link latent infections to RNA viral loads through the shedding kernel \(\{w^{\mathrm{shed}}_{d}\}_{d=0}^{D}\), which encodes the timing of viral release after infection. Using \eqref{eq:w_shed}, the expected load is
\[
\pi_{t} = \beta \sum_{d=0}^{D} I_{t-d} \, w^{\mathrm{shed}}_{d},
\]
where \(\beta>0\) converts infections into billions of genome copies per day per infection.

The viral loads are modeled by a log-normal likelihood:
\[
\mu_{t}  =  \log \bigl(\pi_{t}\bigr)  -  \tfrac12 \sigma_{y}^{2},
\qquad
y_{t} \sim \mathrm{LogNormal} \bigl(\mu_{t}, \sigma_{y}^{2}\bigr),
\]
where $\sigma_{y}>0$ captures measurement variability on the log scale. This parametrization implies $\mathbb{E}[ y_{t}\mid \pi_{t} ] = \pi_{t}$ and is standard in wastewater applications with multiplicative experimental and sampling variability.

\subsection{State-space model}

We also obtain filtered trajectories from a non-linear Gaussian state–space model (SSM) and propagate their time‑varying variances as known observation uncertainty. This alternative input is used to check whether inferences are sensitive to prefiltering, while allowing week‑specific variances to be carried into the renewal likelihood.

Our methodologies account for noise and missing data in the wastewater time series while separating persistent structure from high‑frequency variability. We implement this step with the \texttt{MARSS} package in \textsf{R} \citep{holmes2012marss}.

Following the approach in \citep{Ensor2025}, let $x_t$ denote the latent wastewater RNA viral load on a log-scale at time $t$ with initial state $x_0 \sim \mathcal{N}(\psi,1)$ where $\psi$ is unknown. The state equation is the first difference twice or 
\[
x_t = 2x_{t-1} - x_{t-2} + w_t, 
\quad w_t \sim \mathcal{N}(0,\sigma_w^2),
\]
where $w_t$ is a Gaussian innovation with mean zero and variance $\sigma_w^2$.

The observation equation is given by
\[
y_t = x_t + v_t,
\quad v_t \sim \mathcal{N}(0,\sigma_v^2),
\]
where $y_t$ denotes the measured viral load on a log scale in week $t$, and $v_t$ represents measurement error; assumed Gaussian with mean zero and variance $\sigma_v^2$.

The parameters of the SSM namely $\sigma_v$, $\sigma_w$, and $\psi$ via maximum likelihood, yielding filtered estimates of the latent trend and its pointwise uncertainty:

\[
\widehat{x}_{t|t} = \mathrm{E}\left[x_{t}|y_{1:t}\right], \quad \widehat{P}_{t|t} = \mathrm{Var}\left[x_t \mid y_{1:t}\right].
\]

These variances are later used to construct uncertainty bands and to propagate measurement uncertainty into subsequent modeling layers. Importantly, estimation is performed on the log-10 scale and subsequently converted to the natural measurement scale (gc/L), ensuring consistency with the log-normal likelihood adopted in the measurement layer of the renewal model.  

Finally, to assess robustness to measurement noise, we first fit the renewal model to the viral-load measurements $y_t$ while injecting week-specific observation variances from the Kalman filter. We then replace $y_t$ with the SSM filtered trajectories $\widehat{y}_t$, again using $\widehat{P}_t$ as known variances. As shown in Section~\ref{sec:results}, the resulting trajectories are practically equivalent; accordingly, we present the SSM-filtered input as the default and treat the direct-measurement fit as a robustness check.

We estimated the parameters of the Bayesian renewal model in \texttt{nimble} via MCMC. The sampler jointly sampled the latent trajectories and the key parameters, producing posterior draws for all unknown quantities. We run multiple chains and discard the warm-up iterations. We assessed convergence using the rank-normalized split $\widehat{R}$ and verified well-mixed chains using the relative Monte Carlo standard error (relMCSE). Finally, for the state-space model, we performed the estimations using maximum likelihood with the BFGS as our optimization option in the \texttt{MARSS} package, which utilizes Kalman filtering for efficient likelihood evaluation.

In the Bayesian layer of the renewal model, the latent driver $z_t$ follows a Gaussian random walk with a diffuse normal prior for $z_1$ and a log-normal prior on the innovation scale $\sigma_{\epsilon}$. The softplus curvature parameter $k$ is assigned a truncated log-normal prior to ensure positivity and a stable mapping between $z_t$ and $R_t$. When $\beta$ is estimated, we place a normal prior on $\log \beta$, inducing a log-normal prior on the natural scale. The observation standard deviation, $\sigma_{y}$, is either supplied externally as a time-varying sequence or treated as a single unknown parameter with a truncated normal prior. All priors are moderately informative, providing reasonable regularization while allowing the likelihood to dominate the inference. The mean of the initial state $\psi$ in the state-space model is estimated separately via maximum likelihood.

\section{Results}\label{sec:results}

In this section, we present results from applying the Bayesian renewal model to RSV data from the WWTP under two specifications: 
(A) viral-load measurements transformed to billions of copies per day (BCPD) with a homoscedastic observation variance $\sigma_y$ estimated; and
(B) state-space–model (SSM) filtered trajectories $\widehat{y}_t$ used as the observation input, with week-specific variances $\widehat{P}_t$ treated as known.

Results are organized as follows:
\begin{enumerate}
  \item \textbf{Convergence diagnostics and parameter estimates}, summarizing stability and consistency across fits; and
  \item \textbf{Latent trajectories of incidence ($I_t$) and effective reproduction number ($R_t$)}, comparing the raw viral-load input versus the filtered input.
\end{enumerate}

\subsection{Diagnostics}

Table \ref{tab:post+conv_combined} summarizes posterior estimates and convergence diagnostics for the scalar parameters under two specifications: (A) viral-load measurements as input with homoscedastic noise (a single $\sigma_y$); and (B) SSM-filtered estimates of the viral load with heteroscedastic noise ($\widehat P_t$). Convergence is strong in both cases: rank-normalized split $\widehat R$ values lie between $1.003$ and $1.018$, and relative Monte Carlo standard errors (relMCSE) are small ($\approx 0.002$–$0.023$), indicating well-mixed chains for all parameters. The link-curvature parameter $k$ centers around $8$ with $\widehat R \le 1.007$ (relMCSE $\le 0.009$), consistent with a moderately sharp softplus mapping from $z_t$ to $R_t$. The innovation standard deviation $\sigma_\epsilon$ is small in both specifications (A: $0.073$; B: $0.078$), suggesting gradual week-to-week changes in transmission. In specification A, dispersion is summarized by $\widehat\sigma_y \approx 0.385$ ($\widehat R=1.003$, relMCSE $=0.002$), whereas in specification B the observation heteroscedasticity is encoded via the time-varying $\widehat P_t$. Overall, posterior summaries are broadly similar across specifications: point estimates differ slightly, but credible intervals overlap and convergence metrics are comparable.

For the plant-specific scale, $\beta$ is $13.827$ (SD $3.646$; 95\% CI $[7.838,\,22.273]$) in A and $14.737$ (SD $3.969$; 95\% CI $[8.591,\,23.945]$) in B; the credible intervals overlap, indicating \emph{no practical difference} in scale across the two specifications. All parameters are well identified, and the diagnostics ($\widehat{R}$ and relMCSE) show comparable sampling performance in A–B. Conceptually, using filtered inputs with time-varying variances $\widehat P_t$ provides an alternative to the homoscedastic model without changing the qualitative conclusions about transmission dynamics.

\begin{table}[htbp]
\centering
\small
\setlength{\tabcolsep}{6pt}
\renewcommand{\arraystretch}{1.15}
\begin{tabular}{lcccc}
\toprule
& \multicolumn{4}{c}{\textbf{A}}\\
\cmidrule(lr){2-5}
& \textbf{Mean} & \textbf{SD} & \textbf{95\% CI} & $\widehat{R}$/relMCSE \\
\midrule
$\beta$             & 13.827 & 3.646 & {\small [7.838, 22.273]} & 1.008 / 0.020 \\
$k$                 & 8.238  & 4.691 & {\small [2.122, 18.681]} & 1.007 / 0.008 \\
$\sigma_{\epsilon}$ & 0.073  & 0.028 & {\small [0.030, 0.135]}  & 1.008 / 0.023 \\
$\sigma_{y}$        & 0.385  & 0.038 & {\small [0.316, 0.466]}  & 1.003 / 0.002 \\
\midrule
& \multicolumn{4}{c}{\textbf{B}}\\
\cmidrule(lr){2-5}
& \textbf{Mean} & \textbf{SD} & \textbf{95\% CI} & $\widehat{R}$/relMCSE \\
\midrule
$\beta$             & 14.737 & 3.969 & {\small [8.591, 23.945]} & 1.018 / 0.021 \\
$k$                 & 8.026  & 4.727 & {\small [1.765, 18.600]} & 1.005 / 0.009 \\
$\sigma_{\epsilon}$ & 0.078  & 0.027 & {\small [0.032, 0.137]}  & 1.017 / 0.019 \\
$\sigma_{y}$        & ---    & ---   & ---                        & --- \\
\bottomrule
\end{tabular}
\caption{Posterior means, standard deviations, 95\% credible intervals, and convergence diagnostics ($\widehat R$/relMCSE) for scalar parameters under two specifications: (A) BCPD with homoscedastic $\sigma_y$ estimated; (B) SSM-filtered viral-load inputs with week-specific variances $\widehat P_t$ treated as known.}
\label{tab:post+conv_combined}
\end{table}

\subsection{Latent infections and transmission dynamics from wastewater}

\begin{figure}[htbp]
\centering
\includegraphics[width=0.9\textwidth]{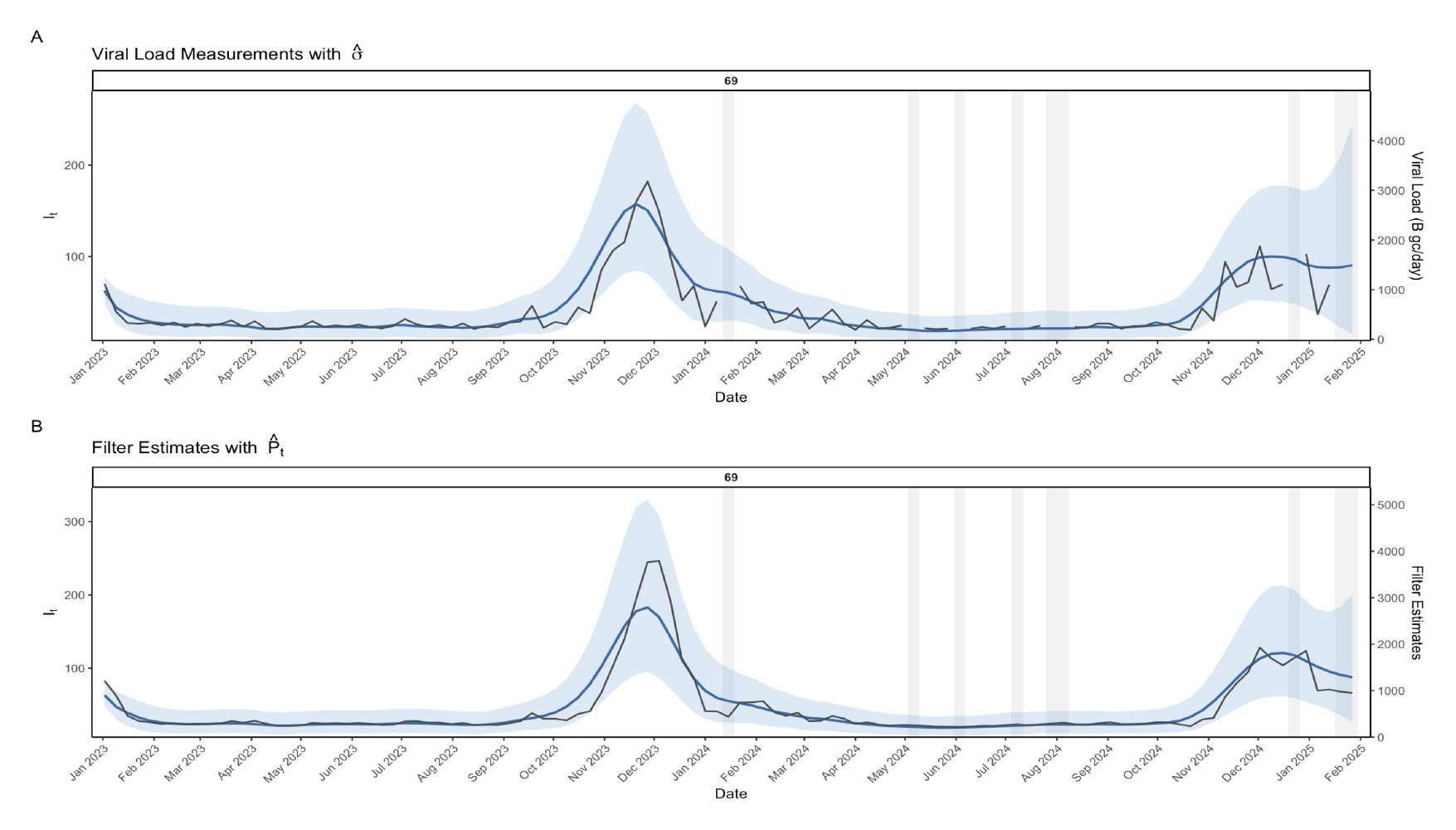}
\caption{Latent incidence ($I_t$) with viral load on the right axis.
Blue line: posterior mean of $I_t$; blue band: 95\% credible interval; gray line: viral load (B gc/day).
Panels A–B correspond, respectively, to:
(A) BCPD with homoscedastic observation variance $\sigma_y$ estimated; and
(B) SSM-filtered trajectories with week-specific variances $\widehat P_t$ treated as known.
Gray vertical bands represent periods with missing wastewater measurements.}
\label{fig:It_panels_69}
\end{figure}

Figure~\ref{fig:It_panels_69} displays posterior means and 95\% credible bands for latent incidence ($I_t$) overlaid with wastewater viral load. Across both specifications (A–B), $I_t$ rises sharply in late 2023, drops in January 2024, remains low through mid-2024, and shows a moderate uptick in autumn 2024. Timing is coherent: the $I_t$ peak precedes the load maximum by roughly one week in each case. Phase, peak timing, and overall $I_t$ trajectories are practically indistinguishable across inputs. Because the absolute scale of $I_t$ is not identified from wastewater alone, we interpret $I_t$ in relative terms and anchor its level via the initial condition.

\begin{figure}[H]
\centering
\includegraphics[width=0.9\textwidth]{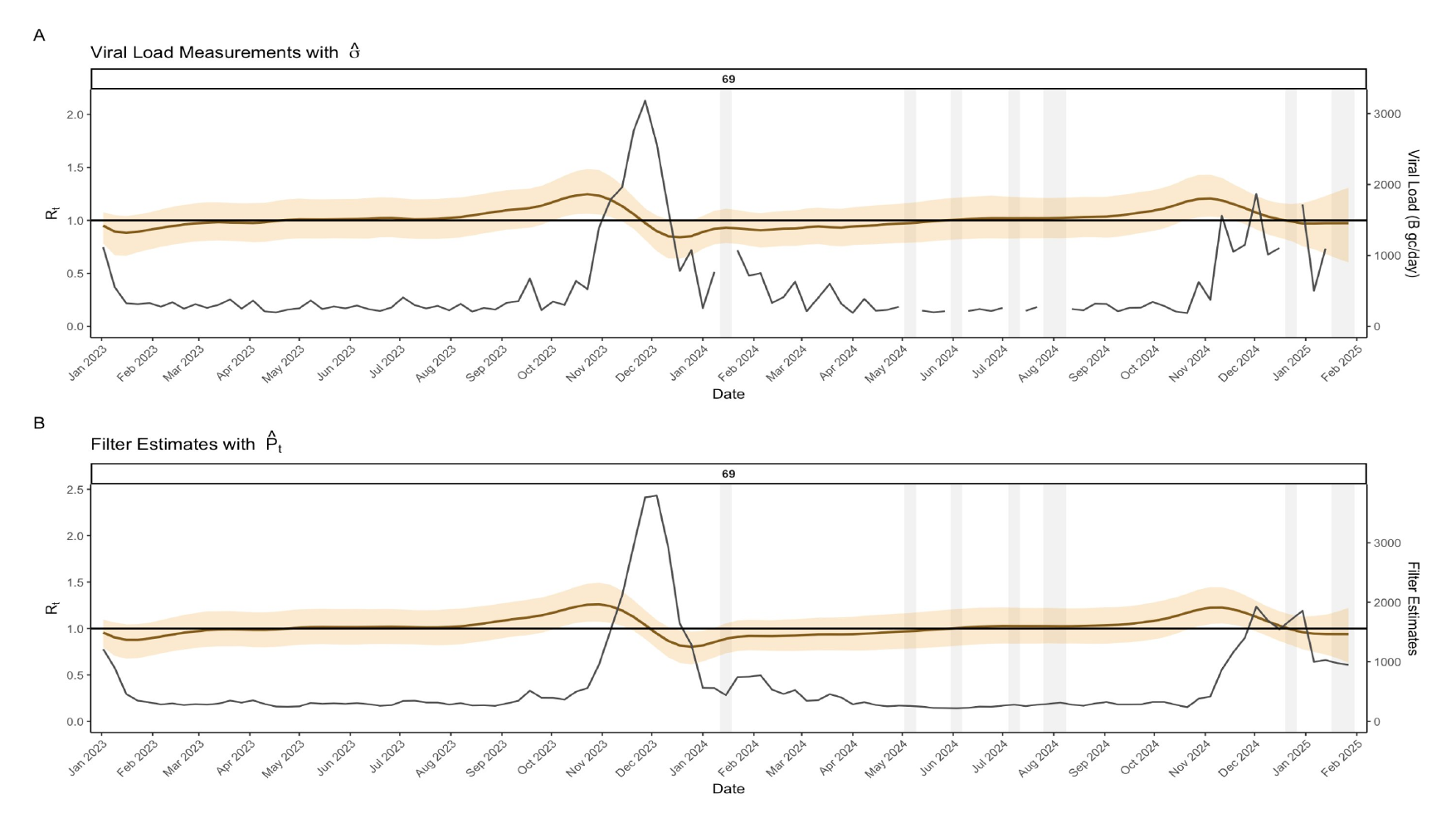}
\caption{Effective reproduction number ($R_t$) with viral load on the right axis.
Orange line: posterior mean of $R_t$; orange band: 95\% credible interval; horizontal line at $R_t=1$; gray line: viral load (B gc/day).
Panels A–B correspond, respectively, to:
(A) BCPD with homoscedastic observation variance $\sigma_y$ estimated; and
(B) SSM-filtered trajectories with week-specific variances $\widehat P_t$ treated as known.
Gray vertical bands represent periods with missing wastewater measurements.}
\label{fig:Rt_panels_69}
\end{figure}

Figure~\ref{fig:Rt_panels_69} reports the effective reproduction number ($R_t$). Across both specifications (A–B), $R_t$ is subcritical ($<1$) in early 2023, becomes clearly supercritical in autumn 2023, and drops below 1 around December–January; a smaller episode appears in autumn 2024. For 69th Street (October 2023–February 2024), the peaks occur on October 23 and 30 for $R_t$, November 20 and 27 for $I_t$, and November 27 and December 4 for the load (top and bottom panels, respectively). In both specifications, the $R_t$ maximum leads the load maximum by about five weeks. The ordering $R_t \rightarrow I_t \rightarrow$ load holds and is consistent with the generation and shedding kernels (about 1–2 weeks) and weekly aggregation. The inferred $R_t$ trajectories are practically indistinguishable across the two modeling specifications.

\begin{figure}[htbp]
\centering
\includegraphics[width=0.9\textwidth]{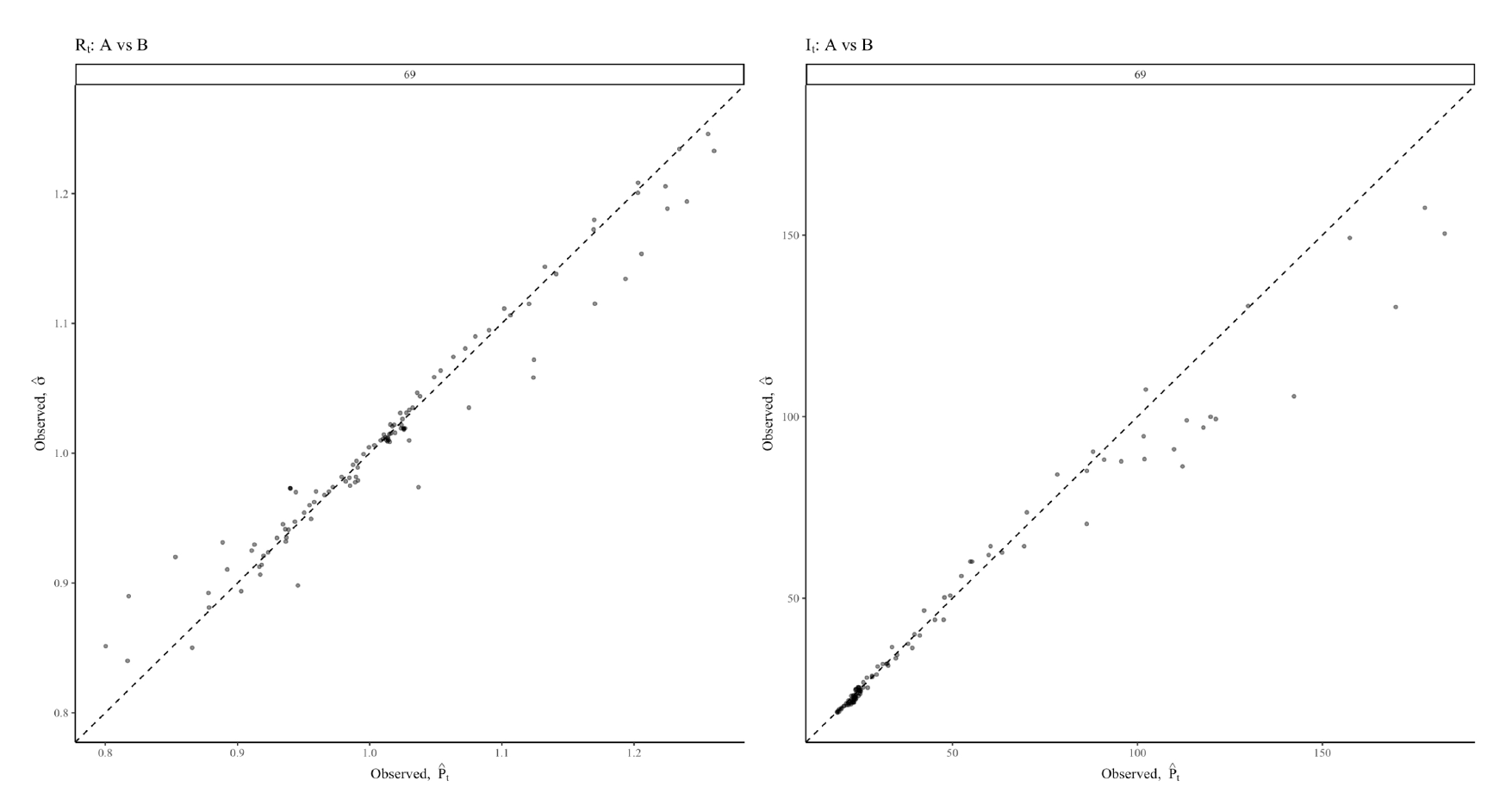}
\caption{Comparison of weekly posterior means under specifications A (BCPD, homoscedastic $\sigma_y$) and B (SSM-filtered inputs with week-specific variances $\widehat P_t$): left, $R_t$; right, $I_t$. Points are weekly posterior means; the dashed line is the identity $y=x$.}
\label{fig:scatter_rt_69}
\end{figure}

To assess practical differences across specifications, Figure~\ref{fig:scatter_rt_69} contrasts weekly posterior means for $R_t$ and $I_t$ under (A) BCPD with homoscedastic noise and (B) SSM-filtered inputs with week-specific variances $\widehat P_t$. For $R_t$, points fall close to the identity with minimal dispersion, indicating that estimates are largely insensitive to using filtered inputs with heteroscedastic variances. For $I_t$, dispersion is slightly larger—most visibly near peak weeks—but without a consistent bias. Overall, systematic differences are negligible and the inferred transmission trajectories are practically indistinguishable across the two specifications.

\section{Discussion}

We used a renewal framework with biologically motivated generation and shedding kernels to infer RSV transmission from wastewater. In contrast to more elaborate pipelines (e.g., \texttt{EpiSewer}) that couple multiple modules and assumptions, and \texttt{ERN}, which adds structural complexity to estimate $R_t$, our approach is intentionally parsimonious and transparent. Latent infections and observed load are linked through two kernels and a small set of identifiable parameters.

We evaluated two input/noise specifications that differ only in how the signal and its uncertainty enter the likelihood: (A) viral‐load measurements in billions of copies per day (BCPD) with homoscedastic noise; and (B) SSM‐filtered load estimates (FE) with week‐specific Kalman variances $\widehat P_t$ treated as known. Convergence diagnostics were satisfactory in both cases (acceptable $\widehat R$ and low relative MCSE). Model comparison and pairwise scatter plots show that $R_t$ is essentially invariant across A–B. For $I_t$, using FE with $\widehat P_t$ (A vs. B) increases dispersion primarily in the upper tail; outside peak periods, trajectories are effectively equivalent. Given this pattern, the filtered input with $\widehat P_t$ is attractive for routine reporting—because it carries week‐specific observation variance—without altering the qualitative epidemiological interpretation.

Timing is consistent across the model specifications. The $R_t$ maximum leads the wastewater maximum by about five weeks, and the $I_t$ peak precedes the wastewater maximum by approximately one week. This ordering aligns with the generation and shedding kernels (on the order of 1–2 weeks) and the weekly aggregation of the series, supporting the renewal-based interpretation.

We also refit the model using concentration in copies per liter rather than flow-normalized copies per day. Dimensionless parameters were essentially unchanged, whereas the plant-specific scale $\beta$ shifted to absorb the unit change. Because switching units multiplies the observed signal by a constant, the model compensates by rescaling $\beta$ while leaving shape and timing intact. In practice, the coefficients are scale-invariant except for $\beta$; the inferred $R_t$ trajectory is unaffected, and $I_t$ changes only by a constant multiplicative factor tied to the measurement scale.

The absolute scale of $I_t$ is not identified from wastewater alone, so we anchor it with $I_{1}\sim\mathrm{Poisson}(61)$, motivated by the fraction of Houston’s population served by the selected sewershed and the city-wide mean weekly RSV cases. Sensitivity checks using alternative anchors for the initial condition confirm that peak timing and the ordering $R_t \rightarrow I_t \rightarrow$ load are preserved and that the $R_t$ trajectory is robust; as expected, the level of $I_t$ shifts nearly proportionally to the chosen anchor with compensating adjustments in $\beta$. Accordingly, we interpret $I_t$ in relative terms.

\section{Conclusions}

Using a parsimonious renewal framework with generation and shedding kernels, we inferred RSV transmission from wastewater. We obtained the relative number of infections and the effective reproduction number with coherent timing: $R_t$ leads $I_t$, and $I_t$ leads the wastewater signal. Comparing two input choices—(A) measurements in billions of copies per day with homoscedastic noise; and (B) SSM-filtered load estimates with week-specific variances from the Kalman filter ($\widehat P_t$)—we find $R_t$ to be essentially invariant across specifications. For $I_t$, using filtered inputs with $\widehat P_t$ (A vs.\ B) increases dispersion primarily in the upper tail; outside peak periods, trajectories are effectively equivalent. Results are also robust to the measurement scale (gc/L vs.\ billions of copies per day): parameters are effectively unchanged except for the plant-specific scaling factor $\beta$, which absorbs unit changes. Overall, wastewater data coupled with a simple renewal structure provide stable estimates of $I_t$ (in relative terms) and $R_t$ without the need for complex preprocessing. When available, we therefore recommend using the SSM-filtered input as the default, with direct measurements retained as a robustness check.

\section*{Funding and Acknowledgments}

The authors disclosed receipt of the following financial support for the research, authorship, and/or publication of this article: This work was supported by the Centers for Disease Control and Prevention (ELC-ED grant no. 6NU50CK000557-01-05 and ELC-CORE grant no. NU50CK000557). The authors acknowledge Houston Public Works for their contributions to the HHD WBE system. The authors would like to acknowledge the CDC National Wastewater Surveillance System (NWSS) scientific community. For more information on the Houston Wastewater Epidemiology Center of Excellence see \url{https://www.hou-wastewater-epi.org}.

\section*{Notes on contributor(s)}

Palacio is the corresponding author, representing both intellectual leadership and implementation of the methodologies. He presented the team with a complete first draft for editorial comments. Ensor, Rice Co-PI of this project, led the technical development and draft writing. Keller consulted on developing the two-stage implementation with specific attention to uncertainty quantification. Schneider oversees the day-to-day implementation of the HHD WBE system and represents the team in national data analysis conversations. Domakonda serves as HHD manager for this project. Hopkins, serves as HHD PI for this project.  Stadler, Rice Co-PI of this project, oversees all aspects of laboratory analyses, and brings the essential expertise in WBE. All authors contributed to the scientific discussion and edited the manuscript.

\bibliographystyle{plainnat}
\bibliography{Epi}

\newpage
\section*{Appendix}

\subsection*{1 Introduction}
This appendix presents two scenarios that complement the main manuscript and use the same modeling specifications A–B:
(A) BCPD with homoscedastic observation variance estimated; and
(B) SSM-filtered trajectories with week-specific variances $\widehat P_t$ treated as known.

We re–fit the model using \emph{copies per liter} (gc/L) instead of \emph{billions of copies per day} (BCPD). This example isolates the role of units and flow normalization. As shown below, dimensionless parameters and inferred trajectories ($R_t$, $I_t$) are effectively unchanged, while the plant–specific scale $\beta$ adjusts to absorb the unit change.

\subsection*{2 Analysis using gc/L}

\begin{figure}[htbp]
\centering
\includegraphics[width=0.9\textwidth]{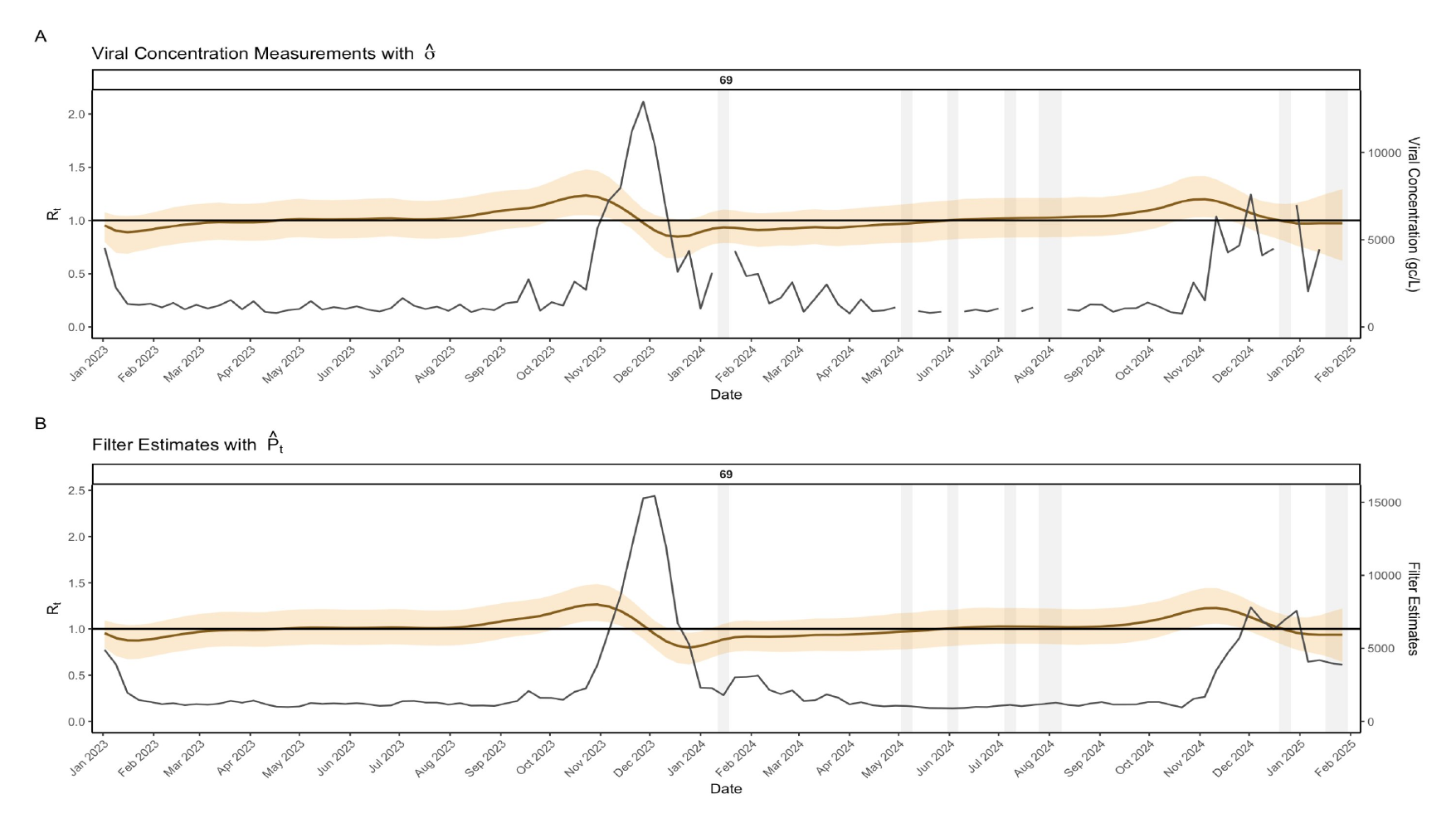}
\caption{Effective reproduction number ($R_t$) with viral load on the right axis.
Orange line: posterior mean of $R_t$; orange band: 95\% credible interval; horizontal line at $R_t=1$; gray line: viral concentration (gc/L).
Panels A–B correspond, respectively, to:
(A) viral concentrations with homoscedastic observation variance $\sigma_y$ estimated; and
(B) SSM-filtered trajectories with week-specific variances $\widehat P_t$ treated as known.}
\label{fig:Rt_panels}
\end{figure}

\begin{figure}[htbp]
  \centering
  \includegraphics[width=0.9\textwidth]{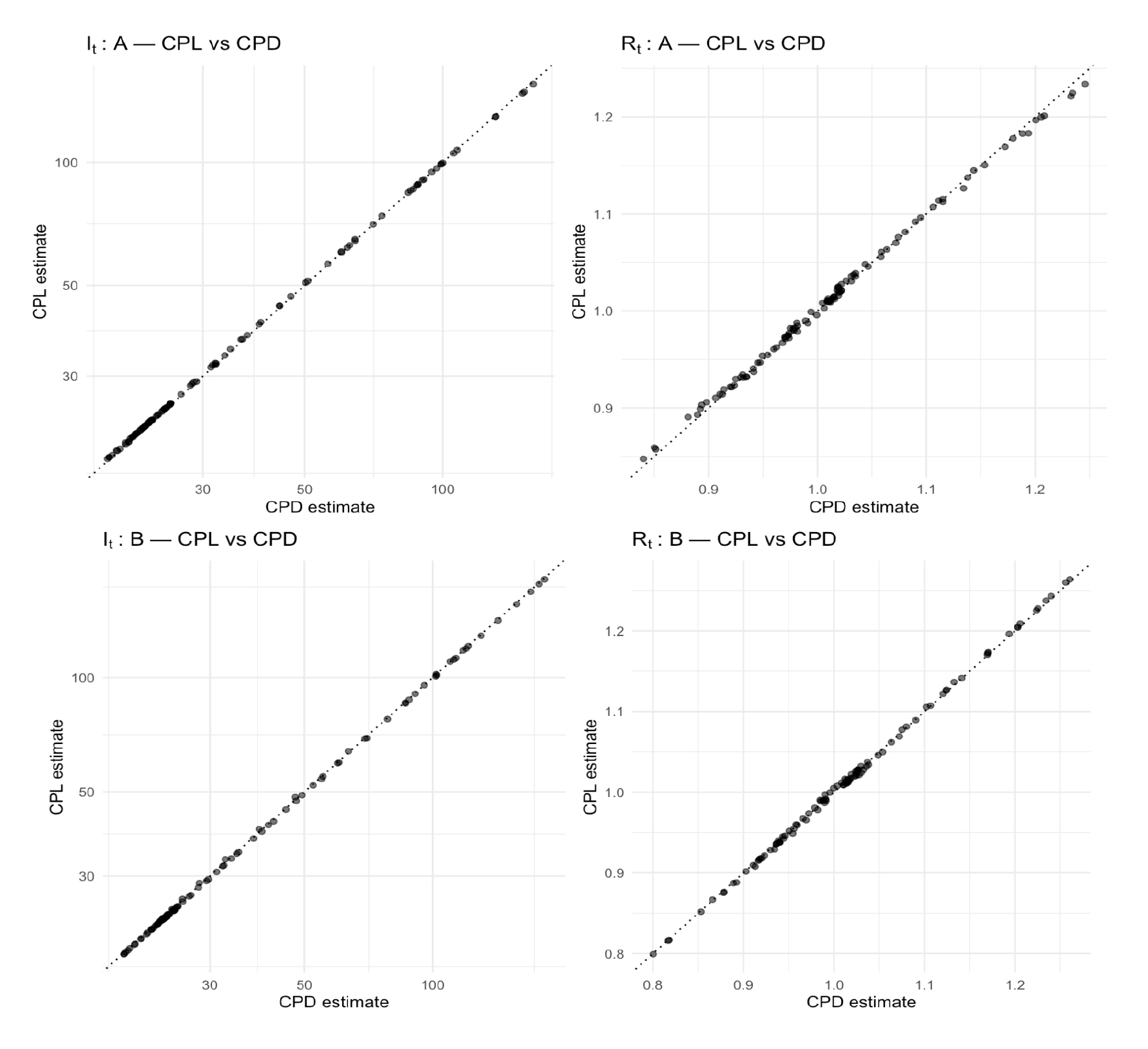}
  \caption{Pairwise scatterplots comparing estimates from viral concentration (CPL) and viral load (CPD) for $I_t$ (left) and $R_t$ (right) under models A–B. Points are weekly posterior means; the dashed line is $y=x$.}
  \label{fig:scatter_conc_load_69}
\end{figure}

For $R_t$, the viral concentration panels in Fig.~\ref{fig:Rt_panels} sit essentially on top of their counterparts based on viral load: same autumn-2023 maximum, the dip around December–January, and the minor late-2024 rise. The concentration–load scatterplots in Fig.~\ref{fig:scatter_conc_load_69} reinforce this invariance—points hug the identity with correlations of $0.999$ and very small MAE ($0.002$–$0.003$; Table~\ref{tab:agreement_cpl_cpd}). Parameter summaries are consistent with this picture: the dynamic/shape parameters ($k$, $\sigma_\epsilon$) remain stable across A–B and across data types (Table~\ref{tab:params_concentration_69}).  In short, $R_t$ is unit-invariant: switching from viral load to viral concentration leaves the epidemiological signal unchanged.

\begin{figure}[htbp]
\centering
\includegraphics[width=0.9\textwidth]{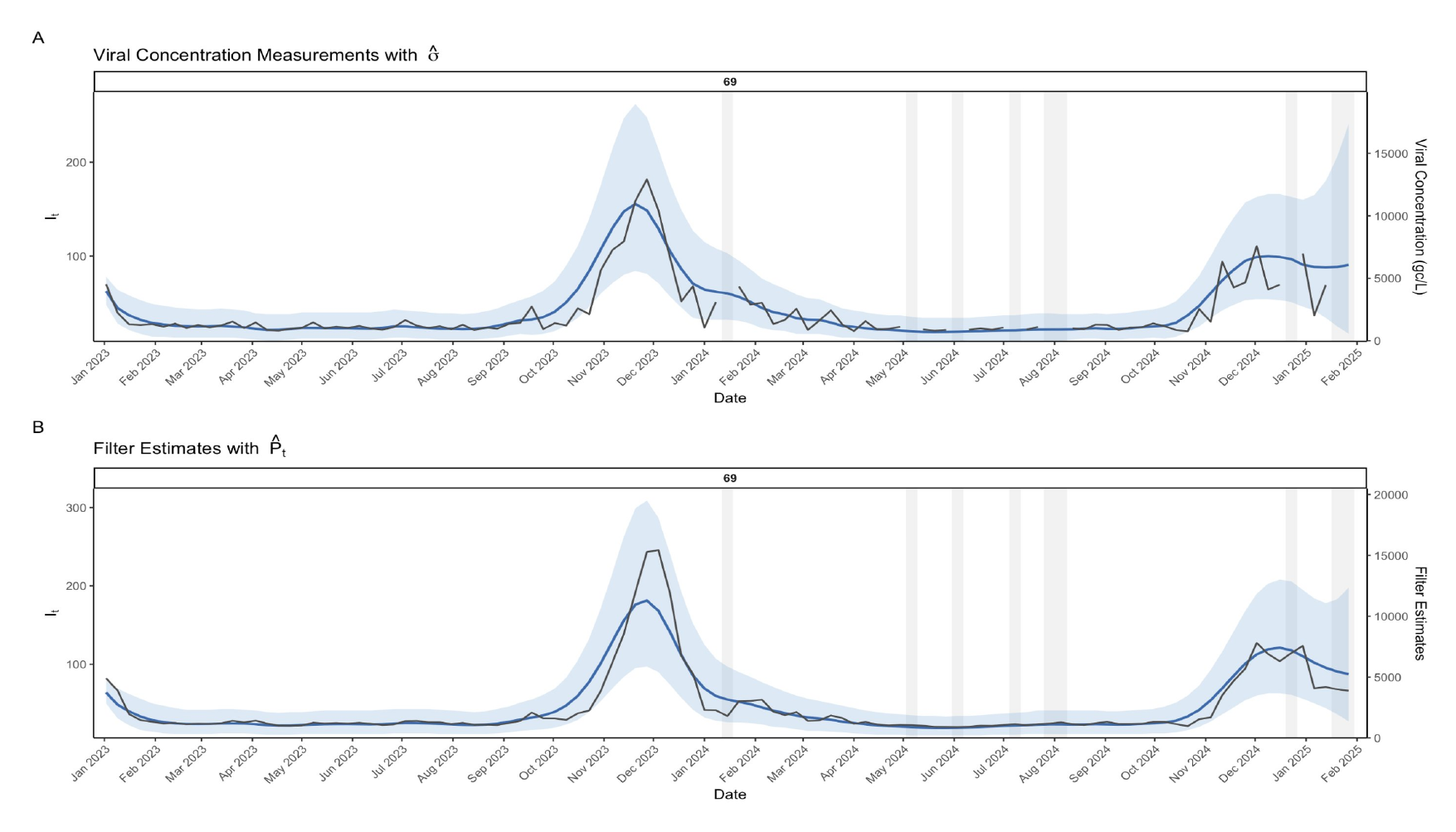}
\caption{Latent incidence ($I_t$) with viral load on the right axis. 
Blue line: posterior mean of $I_t$; blue band: 95\% credible interval; gray line: viral concentration (gc/L). 
Panels A–B correspond, respectively, to:
(A) viral concentrations with homoscedastic observation variance $\sigma_y$ estimated; and
(B) SSM-filtered trajectories with week-specific variances $\widehat P_t$ treated as known.}
\label{fig:It_panels}
\end{figure}

For $I_t$, the viral concentration panels in Fig.~\ref{fig:It_panels} reproduce the same seasonality and peak timing as viral load (late-2023 peak, January decline, mild autumn-2024 uptick). Concentration-based trajectories track load-based ones closely, with only small differences in magnitude during the peak and the late-2024 uptick. These differences are consistent with the modest MAE values for $I_t$ ($0.287$–$3.364$ infections; Table~\ref{tab:agreement_cpl_cpd}) and reflect the expected $\beta$–$I_t$ scale trade-off once flow is removed from the observation map. Timing is preserved and level shifts are minor, indicating that changes in data type affect units rather than the underlying epidemic pattern.

\begin{table}[htbp]
\centering
\small
\setlength{\tabcolsep}{6pt}
\renewcommand{\arraystretch}{1.15}
\begin{tabular}{lccc}
\toprule
\textbf{Metric} & \textbf{Model} & \textbf{Correlation} & \textbf{MAE} \\
\midrule
$I_t$  & A & 0.999 & 0.287 \\
$R_t$  & A & 0.999 & 0.003 \\
\midrule
$I_t$  & B & 0.999 & 3.364 \\
$R_t$  & B & 0.999 & 0.002 \\
\bottomrule
\end{tabular}
\caption{Agreement between viral concentration and viral load estimates (correlation and MAE) for $I_t$ and $R_t$ across models A–B.}
\label{tab:agreement_cpl_cpd}
\end{table}

Table~\ref{tab:params_concentration_69} shows that $k$ and $\sigma_\epsilon$ are empirically stable across data types. For example, the posterior means of $k$ under \emph{load} (A/B: $8.24/8.03$) and under \emph{concentration} (A/B: $8.48/8.02$) are close, and their 95\% credible intervals strongly overlap (e.g., model A $k$: [2.12, 18.68] in load vs. [2.37, 18.83] in concentration; model A $\sigma_\epsilon$: [0.030, 0.135] vs. [0.025, 0.131]; similarly for model B with $\sigma_\epsilon$: [0.032, 0.137] vs. [0.037, 0.132]). Mean shifts are small relative to posterior spreads, supporting the view that these parameters are invariant to the data type. By design, $\beta$ changes scale and absorbs unit effects. For the observational noise $\sigma_y$, posteriors are also close across data types (e.g., $\sigma_{y,\text{load}}=0.385\,[0.316, 0.466]$ vs. $\sigma_{y,\text{conc}}=0.387\,[0.319, 0.466]$).

\begin{table}[htbp]
\centering
\small
\setlength{\tabcolsep}{6pt}
\renewcommand{\arraystretch}{1.15}
\begin{tabular}{lcccc}
\toprule
\textbf{Parameter} & \textbf{Model} & \textbf{Mean} & \textbf{St.Dev.} & \textbf{95\% CI} \\
\midrule
$k$                 & A & 8.483 & 4.685 & {\small [2.367, 18.829]} \\
$\sigma_{\epsilon}$ & A & 0.069 & 0.027 & {\small [0.025, 0.131]} \\
$\beta$             & A & 54.847 & 12.572 & {\small [34.224, 82.537]} \\
$\sigma_{y}$        & A & 0.387 & 0.037 & {\small [0.319, 0.466]}\\ 
\midrule
$k$                 & B & 8.018 & 4.680 & {\small [1.974, 18.596]} \\
$\sigma_{\epsilon}$ & B & 0.079 & 0.024 & {\small [0.037, 0.132]} \\
$\beta$             & B & 59.988 & 15.416 & {\small [36.055, 96.459]} \\
\bottomrule
\end{tabular}
\caption{Posterior summaries for models A–B using viral concentration (gc/L).}
\label{tab:params_concentration_69}
\end{table}
\end{document}